# Scattering Suppression from Arbitrary Objects in Spatially-Dispersive Layered Metamaterials

Alexander S. Shalin[1,3,4,*], Pavel Ginzburg[2,**], Alexey A. Orlov[1], Ivan Iorsh[1], Pavel A. Belov[1], Yuri S. Kivshar[1,5], and Anatoly V. Zayats[2]

[1]ITMO University, St. Petersburg197101, Russia
[2]Department of Physics, King's College London, Strand, London WC2R 2LS, United Kingdom
[3]Kotel'nikov Institute of Radio Engineering and Electronics of RAS (Ulyanovsk branch), Ulyanovsk *432011*, Russia
[4]Ulyanovsk State University, Ulyanovsk 432017, Russia
[5]Nonlinear Physics Center, Australian National University, Canberra ACT 0200, Australia

**Abstract:**

Concealing objects by making them invisible to an external electromagnetic probe is coined by the term "cloaking". Cloaking devices, having numerous potential applications, are still face challenges in realization, especially in the visible spectral range. In particular, inherent losses and extreme parameters of metamaterials required for the cloak implementation are the limiting factors. Here, we numerically demonstrate nearly perfect suppression of scattering from arbitrary shaped objects in spatially dispersive metamaterial acting as an alignment-free concealing cover. We consider a realization of a metamaterial as a metal-dielectric multilayer and demonstrate suppression of scattering from an arbitrary object in forward and backward directions with perfectly preserved wavefronts and less than 10% absolute intensity change, despite spatial dispersion effects present in the composite metamaterial. Beyond the usual scattering suppression applications, the proposed configuration may serve as a simple realisation of scattering-free detectors and sensors.

Corresponding author: *Shalin_a@rambler.ru, **pavel.ginzburg@kcl.ac.uk

1. Introduction

Controlling scattering from an object illuminated by an external wave is one of the most prominent important objectives of applied electromagnetism. Various approaches have been developed and employed in antenna engineering for the enhancement of scattering cross-sections and directivities which are the most demanded characteristics [1]. While the main technological effort has been focused at the millimetre- and radio-frequencies ranges where majority of applications are, considerable interest at the optical frequencies has recently emerged. Optical antennas have been already shown to manipulate radiation properties of single emitters as well as to enhance absorption cross-sections [2].

Reduction of scattering cross-sections in a specifically designed material environment leads to reduced detectability of objects, and may, ultimately, result in "invisible" objects if the scattering is absent. The concept of 'cloaking' was introduced in [3,4] and gained considerable attention due to continuous demand to achieve invisibility for radar waves [5] and visible light [6,7,8]. The general approaches for cloaking rely either on transformation optics concepts [3] or conformal optical mapping of complex electromagnetic potentials [4]. While the former generally results in the requirements of highly anisotropic (and sometimes singular) electric and magnetic susceptibilities of a medium of a cloak, the latter approach requires a position-dependent refractive index variation.

One of the main challenges in the development of practical cloaking devices is to minimize demands on permeabilities and anisotropy and reduce inherent material losses of required materials. The so-called 'carpet cloak' has been proposed and implemented to mitigate these factors by imposing certain geometrical restrictions and utilizing quasi-conformal mapping [9,10]. Qualitatively different approach to cloaking utilises epsilon-near-zero (ENZ) metamaterials to suppress a dipolar scattering of a concealed object [11]. ENZ regime, where a real part of the permittivity is close to zero, can be achieved in anisotropic configurations where wave with certain polarization does not have phase advancement. While number of the ENZ metamaterial realisations exist in the radio-frequency range [12,13], in optics such an anisotropic response may be achieved through a metal-dielectric layered structures [14], semiconductor heterostructures [15], or a vertically aligned arrays of nanorods [16]. Recently, an idealised homogeneous uniaxial ENZ material was proposed for partial cloaking [17], but its realization was not specified. In particular, the presence of inherent strong spatial dispersion, associated with realistic plasmonic metamaterial geometry, was not addressed.

In this paper, we demonstrate suppression of scattering from arbitrary shaped and large (not necessarily subwavelength) sized objects placed inside a layered metal-dielectric metamaterial. We show that a metamaterial realization has major influence on the scattering phenomenon since the electromagnetic response of the plasmonic multilayers is substantially affected by spatial dispersion effects [18] but does not impede scattering suppression. Investigation of the exact numerical model, taking into account material losses and finite dimensions of the metamaterial realization, shows the possibility of almost perfect scattering suppression from arbitrary shaped objects with the minimal variations of the phase front of the transmitted/reflected optical wave and small amplitude modifications. The proposed scheme does not require extreme electric and magnetic susceptibilities and can operate in an alignment-free manner. This approach is qualitatively different from other approaches when the incident light does not interact with the cloaked object but bended around it by a material layer [3, 5] or when the scattered field is suppressed with another anti-parallel dipole [11]. In our approach, the light scattering by an object placed inside a metamaterial is strongly anisotropic and suppressed in the forward (transmission) and backward (reflection) directions.

The manuscript is organized as follows. First, various dispersion regimes in layered metamaterial composites are studied. This is followed by investigations of a dipole radiation pattern inside the metamaterial. Subsequently, concealing of objects of various shapes and dimensions is numerically analyzed in two-dimensional geometries. The scattering properties in three-dimensional case are studied in the final section where both numerical modeling and fully analytical formulation of the scattering problem in the dipolar approximation underline advantages and limitations of the proposed approach.

## 2. Results and Discussion

### 2.1 Spatial dispersion in plasmonic multilayer metamaterial

The effective permittivity tensor of metamaterials based on multilayeres can be directly evaluated from their thickness, periodicity, and optical parameters of the constitutive materials [18]. While this approach can predict optical properties of the composites made of low-contrast, positive permittivity layers, it faces severe challenges once negative permittivity materials (plasmonic metals) are involved. Optical properties of such multilayered composites are influenced by guided surface electromagnetic modes on metal-dielectric interfaces, surface plasmon polaritons (SPPs), and spatial dispersion effects become especially significant in the parameters range where the effective medium theory (EMT) predicts vanishing values of permittivity [19].

SPP modes in planar layered composites, as well as their dispersion, can be found via semi-analytical formulation using the transfer matrix method, and the dispersion equation of eigen modes in arbitrary periodic multilayered structures is given by [18]

$$2\cos(\beta(d_d+d_m)) = \left(\cos(\beta_d d_d)\cos(\beta_m d_m) - \frac{\beta_d}{\beta_m}\sin(\beta_d d_d)\sin(\beta_m d_m)\right), \quad (1)$$

where $\beta$ is the Bloch wave number, $d_d$ and $d_m$ are the thicknesses of dielectric and metal layers, respectively, and $\beta_m$ and $\beta_d$ are the transverse parts of the wave vectors in metal and dielectric materials, respectively, $\beta_{m,d} = \sqrt{\varepsilon_{m,d}\left(\frac{\omega}{c}\right)^2 - k_x^2}$, where $k_x$ is the transverse wave vector in the direction parallel to the layers [18]). Solution of Eq. 1 enables to obtain isofrequency contours of the modes, allowed to propagate in the layered medium for the frequency fixed at arbitrary value. Hereafter, generic metamaterial parameters were chosen to be $d_m$=20 nm for Au [20] layer, $d_d$=100 nm for air (note that any dielectric material could be used with proper recalculation of the cloak parameters).

The isofrequency curves of an infinite metal-dielectric multilayer structure for the extraordinary, TM polarised, modes have strong dependence on the wavelength, and different characteristic dispersions appear in Fig. 1 (a,c,e). One of the distinctive features of these plots is the simultaneous emergence of multiple bands at the given frequency. In other words, for certain values of $k_y$, there exist 2 nontrivial solutions for $k_x$. This effect is the manifestation of strong spatial dispersion, inherently attributed to plasmonic excitations. The transition between various dispersion regimes occurs in the vicinity of the ENZ point, predicted by EMT calculations (equal to 540 nm for considered multilayer parameters). According to EMT, isofrequency surfaces at the wavelengths longer than the ENZ wavelength have pure hyperbolic shape (Fig. 1(a)), while at slightly shorter wavelengths the dispersion is purely elliptic (Fig. 1(e)).

At the ENZ wavelength, the EMT-obtained isofrequency contours are strongly prolate spheroids (Fig. 1(c)) due to the finite losses in the metal layers (in the lossless case, it degenerates to a single line). The exact solution of the dispersion relation (Eq. 1) shows that the EMT description fails near the ENZ frequency. First, the finite period of the structure imposes limitations on the range of **k**-vectors, in the direction normal to the layers. Thus, the hyperbolic shape of the dispersion will deviate close to the edges of the Brillouin zone, $k_y = \pi/(d_m + d_d)$. Furthermore, due to the presence of SPP modes on the multilayer interfaces, strong inherent spatial dispersion of the metamaterial will result in co-existence of two TM polarised waves. Hence, there are two dispersion branches for the same polarisation, in contrary to conventional media [19, 22]. The mode that follows predictions of the EMT is the main mode, while another is called 'additional'. The behaviour of structural spatial dispersion in layered composites is similar to nanorod metamaterials [21,22]. It should be noted, however, that this type of spatial dispersion has pure 'structural' nature and not related to nonlocalities in material components [23,24].

### 2.2. Dipole radiation in metal-dielectric multilayered metamaterial

Scattering of electromagnetic waves by an arbitrary object can be evaluated using multipolar decomposition of the field taking into account the geometry of the problem [25]. The most significant contribution to the far-field scattering is usually emerges from the dipolar term. Following this rule of thumb, we first study the properties of the scattering by a point dipole inside the multilayer metamaterial, before considering a general case of concealment of an arbitrary object.

It is well known, that the dipole radiation patterns are strongly affected by electromagnetic environment. For example, periodically structured media such as photonic crystals with engineered dispersion may enforce flat fronts of dipole radiation [26]. Furthermore, radiation patterns of an emitter situated inside a homogeneous, but anisotropic media may have quite remarkable shapes. The characteristic radiation pattern of a dipole in free space (a spherical wave front modulated by the cosine of the angle of the dipole axis orientation) becomes cross-shaped if placed in a hyperbolic medium, with the cross opening angle being solely defined by the ratio of the permittivity tensor components [e.g., 27]. The details of this cross-shaped radiation pattern significantly depend on specific metamaterial realization.

We used the scattered-field formulation in the finite element modelling software [28] to simulate the field distributions inside the simulation domain which was surrounded with perfectly-matched layers (the periodic boundary conditions were not used in the simulations, the relation will become clear hereafter). The radiation patterns of a vertically-polarized (y-oriented) dipole, placed inside a multi-layered metamaterial block have qualitatively different behaviours in the hyperbolic, elliptic and ENZ regimes (Fig. 1). While all patterns have a cross-shaped envelope, the wave fronts in the hyperbolic and elliptic regimes are significantly curved, whereas the dispersion near the ENZ frequency maintains the wave front flat. Moreover, the metamaterial was designed to be impedance matched with a free space at normal incidence to minimise reflections on the boundaries: the circle representing the dispersion of light in free space passes the vicinity of the degeneracy point of the metamaterial dispersion for this choice of parameters (Fig. 1(c)). As the result, multiple reflections inside the layered composite are mostly suppressed. The impedance matching design and the flat dipole radiation pattern inside the composite provide the necessary ingredients for maintaining a flat front of the incident wave after interaction with an object and low reflection from the metamaterial itself. The presence of the additional waves in the nonlocal regime inherent to any realistic realization of the metamaterial does not impede flatness of the forward and backward radiated wavefronts.

### 2.3. Scattering suppression from two-dimensional objects

Plane-wave-like radiation pattern of a point dipole directly implies that the major dipolar scattering of an object will not distort the front of an incident wave. In order to verify this, we will first consider scattering from perfect electric conductor (PEC) cylinders of different sizes. The cylinder is placed in the centre of a layered metamaterial block of a finite volume (to simulate a realistic scenario) which is illuminated by a plane wave polarized perpendicular to the layers and with the wavevector along them. It is worth noting, that higher-order multipoles also contribute to the scattering, especially in the case of larger objects. Nevertheless, the scattering suppression still could be observed, as will be confirmed in numerical simulations.

The electric field distributions in and outside the metamaterial when the cylinder placed inside and for the empty metamaterial cover show that the flat wave front of the incident wave is maintained in both cases, while the amplitude slightly decreases due to inherent material losses of the metal layers (Fig. 2(a,b)). Since the metamaterial is impedance matched to free space, as was shown above, reflections from both boundaries are almost completely suppressed. Fig. 2(c) shows the difference of the fields between the case of empty cloak and while cloaking the PEC cylinders of 200 nm diameter, demonstrating nearly perfect cancelation of the scattering from the object in the far-field: the amplitude difference does not exceed 3%. It should be emphasised that the general detectable signature of an object is the wavefront distortion of an incident illuminating wave and/or intensity variations across the wavefront (shadow areas). At the same time, absolute transmitted or reflected intensity is hardly quantifiable without reference measurements. Thus, preserving a phase front and an amplitude front undistorted, the proposed approach is perfect for simple and practical implementation of scattering suppression. The same considerations are applicable for observation in reflection with the impedance-matched metamaterial cover (Fig. 2).

Large PEC cylinders and arbitrary shaped objects have also been tested (Fig. 2(d-g)) as well as strongly absorbing and high-index transparent objects of different shapes (not presented, but showing very similar far-field distributions). Nearly perfect scattering suppression was observed in all scenarios with nearly the same performance (cf. Figs. 2(e) and 2(g)). The amplitude difference observed for a large arbitrary-shaped object is about 7% (Fig. 2(g)), only slightly larger than for the big cylinder, showing about 5% variations (Fig. 2(e)). Furthermore, increasing objects' sizes leads to the increased amplitude difference due to the stronger contribution from the multipole scattering terms. While the above considerations were presented for normally incident collimated beams, we also checked the angular dependence of the scattering. While scattering from the object is suppressed, the same as for normal incidence, the angular dependent impedance matching leads to a strong shadow at larger angles of incidence produced by the metamaterial cover itself.

As an intuitive explanation of the described effect, one can consider that the light scattering by an object inside the metamaterial is strongly anisotropic. In the elliptic and hyperbolic regimes this results to convergent and divergent wave fronts, respectively, thus, leading to "shadows" from the objects located inside metamaterial. In the ENZ regime, the dipole radiation and scattering from the object result in plane wave fronts due to peculiarities of the dispersion. The presence of an additional wave in the realisation of the metamaterial, absent in the homogeneous metamaterial, leads to the reduction of the intensity due to its higher loss, but not destroys plane wave front and cloaking. As a result, the presence of the object under such specific conditions causes only the changes of the transmitted plane wave intensity with the rest of the energy is dissipated due to material loss.

For better understanding of the key benefits of the scattering reduction inside the metamaterial, it is useful to compare a half-transparent glass bock and the one made of the layered metamaterial (our design), assuming that both structures have the same transmission coefficient (less than 100%). Now, a scattering object is placed inside each one of those blocks. The glass cover will not be transparent at the objects' location, whereas the metamaterial composite will maintain the same optical properties, as if no scatterer is present inside. It should also be noted that the importance of the own shadow of the metamaterial cloak on its performance is not unambiguous and strongly depends on the final goal of the application. For perfect invisibility, the shadow should be suppressed; however, even without this, the metamaterial may have possible applications too, such as, e.g., static camouflage cloaking.

### 2.4. Scattering suppression from three-dimensional objects

For three-dimensional objects, the influence of the third dimension introduces additional scattering channels emerging from edge effects of finite length objects placed inside anisotropic metamaterial. In order to estimate the influence of these edge effects, we compared scattering inside the metamaterial from infinite and finite (200 nm length) cylinders. In these simulations, the metamaterial cover was considered with periodic boundary conditions imposed in the direction normal to the layers (y-direction) in order to mitigate the requirements on computational resources.

The scattering amplitude distribution of an infinitely long cylinder (Figs. 3(a)) possesses identical behaviour to the pure 2D case (Fig. 2(d)) confirming that the cloaking effect is not attributed to special properties of Maxwell's equations in low-dimensional space. In the 3D case, the scattering suppression takes place in both x-z and x-y planes (the x-y plane is not shown but identical to the 2D case) as illustrated by the electric field distributions. Consequently, an observer situated outside the metamaterial slab will not detect any wavefront distortions that take place without the cover (amplitude variations are about 5%) as one may see in the inset of Fig. 3. In the case of a finite-sized cylinder (Fig. 3b), while the cloaking in the x-y plane is still preserved (Fig. 3c) with the field distribution and the energy flow lines are almost identical to the 2D case, the wave front in the x-z plane is slightly distorted (Fig. 3d). This is due to the effect of the cylinder edges. The diffraction on the edges gives rise to the appearance of x- and z- components of the scattered field and to the related energy outflow in the direction normal to the propagation direction of the incident wave. It is worth noting that this effect is diminished when long enough (larger than wavelength) particles are considered.

### 2.5. Theory of a three-dimensional homogenised metamaterial cover

In order to understand concealing performance of the metamaterial cover, we have developed a theoretical model treating the scattering process, in the dipolar approximation, in a homogenised anisotropic medium with the permittivity tensor components corresponding to the effective medium permittivity of the metamaterial realisation, described above.

Electromagnetic scattering can be considered in the framework of the Greens functions formulation. We first describe a two-dimensional geometry with infinitely-long cylindrical scatterers situated in a uniaxial anisotropic medium, illuminated perpendicular to its axis with the plane wave having the electric field in y-direction, also perpendicular to the cylinder axis. Using the coordinate transformations, the y-component of the electric field scattered by a 2D electric dipole oriented in y-direction is given by [29]:

$$E_y = p \frac{H_0^{(1)}(k_0 r_e) y^2 \varepsilon_{xx} + x^2 \varepsilon_{yy} \left(k_0 r_e H_0^{(1)}(k_0 r_e) - H_1^{(1)}(k_0 r_e)\right)}{k_0 r_e^3 \sqrt{\varepsilon_{yy}}}, \quad (2)$$

where $r_e = \sqrt{x^2 \varepsilon_{yy} + y^2 \varepsilon_{xx}}$, $p$ is the polarizability of the dipole, $k_0 = \omega/c$, $H_{0,1}^{(1)}$ are the Hankel functions of the first kind, and $\varepsilon_{ii}$ are the diagonal components of the permittivity tensor of the medium. In the limit of $\varepsilon_{xx} \approx 0$, Eq. (2) simplifies to

$$E_y \approx \frac{H_1^{(1)}(k_0 |x| \sqrt{\varepsilon_{yy}})}{k_0 |x| \varepsilon_{yy}} - \frac{H_0^{(1)}(k_0 |x| \sqrt{\varepsilon_{yy}})}{\sqrt{\varepsilon_{yy}}}. \quad (3)$$

Eq. (3) shows that the scattered field has a plane wavefront, the same as the incident field. Therefore, such a configuration provides the low scattering cross-section and the invisibility effect in the far-field observations as the asymptotic Hankel functions for large $|x|$ further simplify Eq. (3) to

$$E_y \approx \frac{i-1}{\sqrt{\varepsilon_{yy}\pi}} \frac{e^{ik_0 x\sqrt{\varepsilon_{yy}}}}{\left(k_0 x\sqrt{\varepsilon_{yy}}\right)^{1/2}}, \qquad (4)$$

corresponding to a unperturbed plane wave.

In the three-dimensional case, the y-component of the electric field scattered at the point 3D dipole aligned along y is in the limit of $\varepsilon_{xx} \approx 0$ is given by [30]:

$$E_y = -ip\frac{1}{\sqrt{\varepsilon_{yy}}}\left(e^{ik_0\sqrt{\varepsilon_{yy}(x^2+y^2)}} - e^{ik_0\sqrt{\varepsilon_{yy}}|x|}\right)\frac{1}{k_0^2 y^2}. \qquad (5)$$

Looking at z=0 plane, one can see that the scattered field is the combination of the wave with a plane wavefront and a cylindrical wave. The cylindrical wave appears due to the excitation of the transverse electric (TE) waves, with the electric field along x-direction, by a 3D point-dipole; the excitation of the TE waves is prohibited in the 2D geometry where translational symmetry along z-axis is imposed. Thus, in 3D case, the cloaking performance in generally reduced for the anisotropic cloak described above, as was also confirmed in Section 2.4 in the simulations of the 3D scattering objects inside the cloak. While the consideration of the composite realization of the metamaterial will introduce nonlocal effects, not considered above for a homogenised cloak, their role will be in the reduction of the overall intensity and not the phase-front distortion, as shown in the previous section.

3. Conclusion

We have shown that the layered plasmonic metamaterial works as nearly-perfect cloaking device for arbitrary-shaped objects of arbitrary sizes. Underlining the major differences between realistic (e.g., layered) composite metamaterial realisation and perfect homogeneous metamaterial [17], we show that the effects of spatial dispersion could have strong influence on the scattering dynamics, nevertheless do not prevent the cloaking phenomenon. The fundamental principle behind this spatially dispersive type of cloaking relies on the coherent scattering of incident waves in the normal and additional modes supported a spatially-dispersive material. As the result, the transmitted field is only slightly attenuated with the wavefront maintaining its original profile, making the detection (almost always based on the observations of amplitude and wave front aberrations) by a prying observer to be impossible. The same considerations are applicable for observation in reflection with the impedance-matched cloak. The nearly perfect operation for 2D objects has been shown to be preserved for relatively long (longer than the wavelength) 3D arbitrary shaped objects. The proposed type of cloaking may find use not only in the conventional concealing applications but also serve as a platform for 'cloaked detectors' schemes [31] where bulk and alignment free cloaks may significantly reduce the complexity of the realisation.

**Acknowledgments.** This work has been supported in part by EPSRC (UK), the ERC iPLASMM project (321268) and the US Army Research Office(Grant No. W911NF-12-1-0533). A.Z. acknowledges support from the Royal Society and the Wolfson Foundation. A.Z. rejects any support from all the grants funded by Russian Federation and listed below. A.S., A.O., P.B., and Yu.K. have been supported by the Ministry of Education and Science of the Russian Federation (Project 11.G34.31.0020),the President of Russian Federation (Grant SP-2154.2012.1), and the Government of Russian Federation (Grant 074-U01). A.S. acknowledges the support of the Ministry of Science and Education of the Russian Federation (GOSZADANIE 2014/190), and Russian Fund for Basic Research within the project N13-02-00623. The work on numerical simulations and investigating of the field distributions has been funded by the Russian Science Foundation Grant No. 14-12-01227.

**Figure Captions**

**Figure 1** (Colour online). Isofrequency contours for the infinite, multilayered metamaterial simulated for (a) the wavelength of 600 nm, (c) the operational wavelength of 540 nm, and (e) 500 nm wavelength. The metamaterial consists of the air and Au layers of $d_a = 100$ nm and $d_g = 20$ nm thickness, respectively. Red curves show propagation modes of the layered metamaterial: (bright) main modes, (pale) additional modes. Green circle represents light cone in air. Yellow curves are the isofrequency contours of the same metamaterial simulated in the EMT. (b,d,f) The electric field (y-component) distributions of the radiating y-polarized dipole situated in the centre of the metamaterial block of 4.8x4µm size at the wavelength corresponding to the isofrequency contours in (a,c,e). Wavenumbers $k_x$ and $k_y$ are normalized to the Brillouin zone boundaries $\pi/(d_a + d_g)$. White lines represent the power flow.

**Figure 2** (Colour online). Electric field (y-component) distributions for the plane wave illuminating (a) the empty cloak, (b) PEC cylinder having diameter of 200nm placed inside the cloak. Field distribution difference between (a) and (b) is shown in (c). (d) the cloak with the 500 nm diameter PEC cylinder inside and (e) difference between the field distributions in (d) and (a). (f) Electric field (y-component) distribution for the plane wave illuminating the multilayered cloak with the PEC object of an arbitrary shape and larger than the wavelength inside with (g) the field difference compared to the empty cloak. The metamaterial parameters and the wavelength correspond to the dispersions in Fig. 1 (b). White lines represent the power flow. The simulations are performed in 2D geometry.

**Figure 3** (Colour online). Electric field (y-component) distributions for the plane wave illuminating the multilayered cloak in 3D geometry with (a) the infinite length and (b,c,d) finite (200 nm) length cylinder of 500 nm diameter inside: (c) the electric field distribution in x-y plane, (d) the electric field

distribution in x-z plane. Inset shows scattering on the cylinder without the cloak. The metamaterial parameters and the wavelength correspond to the dispersions in Fig. 1 (b).

**Figure 1**

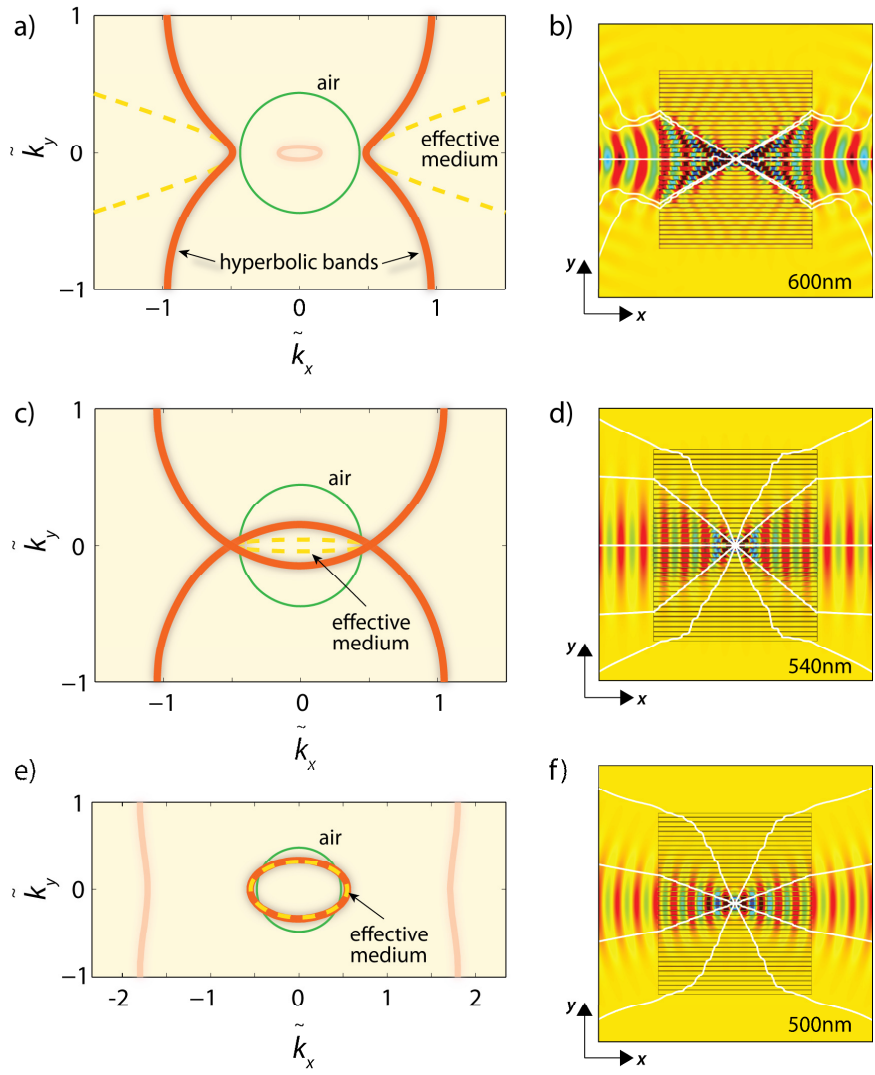

**Figure 2**

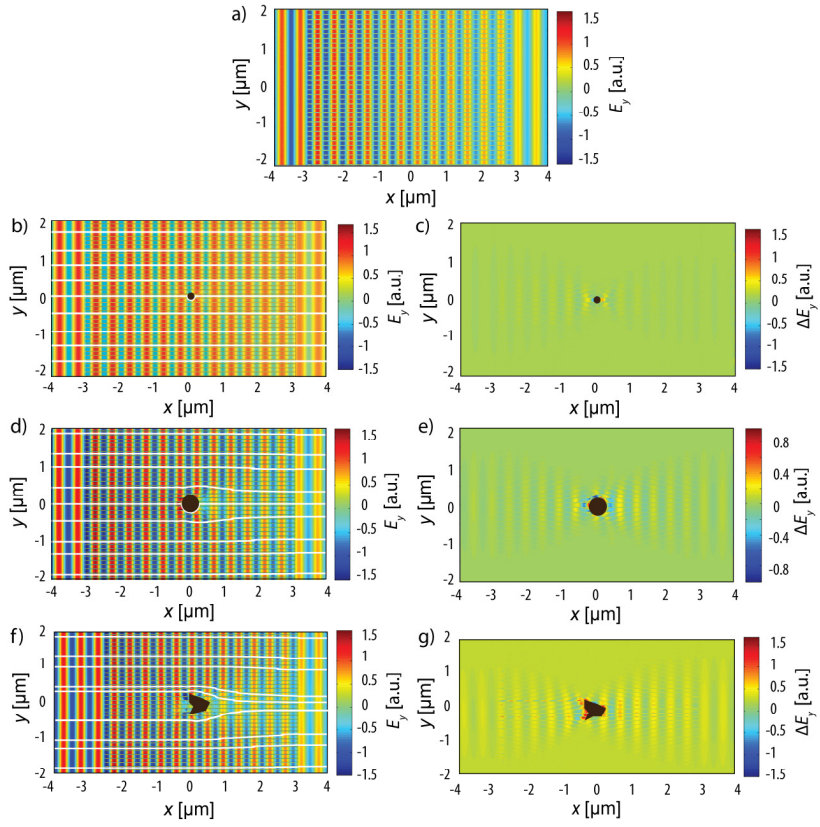

**Figure 3**

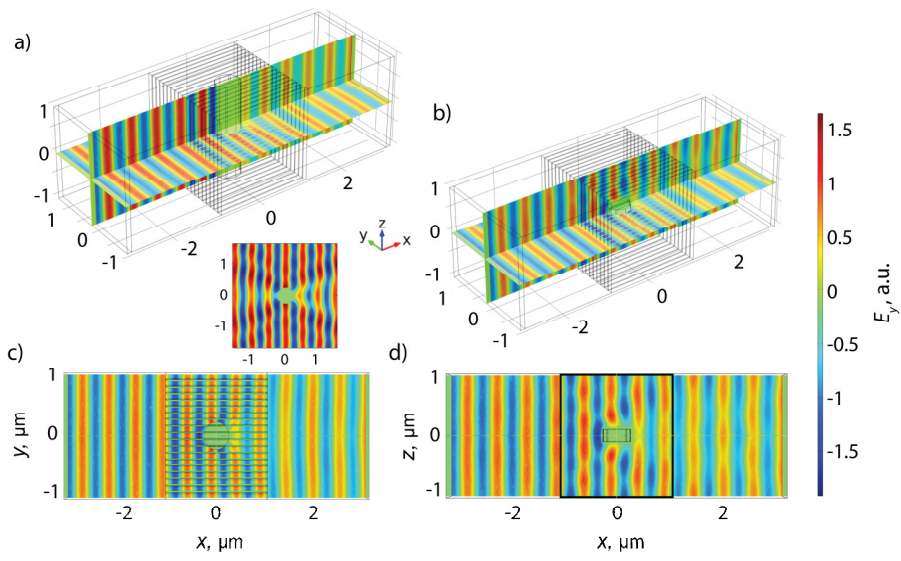

**References:**

1. W. L. Stutzman and G.A. Thiele, *Antenna Theory and Design*, 3rd Ed., John Wiley & Sons, (2012).

2. L.Novotny and N. van Hulst, "Antennas for light", Nature Photon. **5**,83–90, (2011).

3. J. B. Pendry, D. Schurig, and D. R. Smith, "Controlling Electromagnetic Fields", Science Vol. **312** 1780-1782, (2006).

4. U. Leonhardt, "Optical Conformal Mapping", Science **312** 1777-1780, (2006).

5. D. Schurig, J. J. Mock, B. J. Justice, S. A. Cummer, J. B. Pendry, A. F. Starr, and D. R. Smith, "Metamaterial Electromagnetic Cloak at Microwave Frequencies," Science **314**, 977- 980 (2006).

6. J. Valentine, J. Li, T. Zentgraf, G. Bartal, and X. Zhang, "An optical cloak made of dielectrics", Nature Materials **8**, 568 - 571 (2009).

7. L.H. Gabrielli, J. Cardenas, C. B. Poitras, and M. Lipson, "Silicon nanostructure cloak operating at optical frequencies", Nature Photon.**3**, 461 - 463 (2009).

8. W. Cai, U. K. Chettiar, A. V. Kildishev, and V. M. Shalaev, "Optical cloaking with metamaterials", Nature Photon. **1**, 224 - 227 (2007).

9. J. Li and J. B. Pendry, "Hiding under the Carpet: A New Strategy for Cloaking", Phys. Rev. Lett. **101**, 203901 (2008).

10. N. I. Landy and W. J. Padilla, "Guiding light with conformal transformations", Opt. Express **17**, 14872 (2009).

11. A. Alu and N. Engheta, "Achieving Transparency with Metamaterial and Plasmonic Coatings," Phys. Rev. E 72, 16623 (2005).

12. B. Edwards, A. Alù, M. E. Young, M. Silveirinha, and N. Engheta, "Experimental verification of $\varepsilon$-near-zero metamaterial coupling and energy squeezing using a microwave waveguide," Phys. Rev. Lett. **100**, 033903 (2008).

13. R. Liu, Q. Cheng, T. Hand, J. J. Mock, T. J. Cui, S. A. Cummer, and D. R. Smith, "Experimental demonstration of electromagnetic tunneling through an epsilon-near-zero metamaterial at microwave frequencies," Phys. Rev. Lett.**100**, 023903 (2008).

14. Z. Jacob, J. Y. Kim, G. V. Naik, A. Boltasseva, E. E. Narimanov, and V. M. Shalaev, "Engineering photonic density of states using metamaterials," Appl. Phys. B **100**(1), 215–218 (2010).

15. P. Ginzburg, and M. Orenstein, "Nonmetallic left-handed material based on negative-positive anisotropy in low-dimensional quantum structures", J. Appl. Phys. **103**, 083105 (2008).

16. P. Ginzburg, F. J. Rodríguez Fortuño, G. A. Wurtz, W. Dickson, A. Murphy, F. Morgan, R. J. Pollard, I. Iorsh, A. Atrashchenko, P. A. Belov, Y. S. Kivshar, A. Nevet, G. Ankonina, M. Orenstein, and A. V. Zayats, "Manipulating polarization of light with ultrathin epsilon-near-zero metamaterials," Opt. Express **21**, 14907-14917 (2013).

17. J. Luo, W. Lu, Z. Hang, H. Chen, B. Hou, Y.Lai, and C. T. Chan, "Arbitrary Control of Electromagnetic Flux in Inhomogeneous Anisotropic Media with Near-Zero Index", Phys. Rev. Lett. 112, 073903 (2014).

18. A. A. Orlov, S. V. Zhukovsky, I. V. Iorsh, and P. A. Belov, "Controlling light with plasmonic multilayers", Photonics and Nanostructures - Fundamentals and Applications 12 , 213 (2014).

19. A. Orlov, P.Voroshilov, P.Belov, and Y.Kivshar, "Engineered optical nonlocality in nanostructured metamaterials," Phys. Rev. B**84**, 045424 (2011).

20. E. D. Palik, Handbook of Optical Constants of Solids (Academic Press, 1985).

21.R. J. Pollard, A. Murphy, W. R. Hendren, P. R. Evans, R. Atkinson, G. A. Wurtz, A. V. Zayats, and Viktor A. Podolskiy, "Optical Nonlocalities and Additional Waves in Epsilon-Near-Zero Metamaterials", Phys. Rev. Lett. **102**, 127405 (2009).

22.B. M. Wells, A. V. Zayats, and V. A. Podolskiy, "Nonlocal optics of plasmonic nanowire metamaterials", Phys. Rev. B **89**, 035111 (2014).

23.V. M. Ginzburg and V. L. Agranovich, *Spatial Dispersion in Crystal Optics and the Theory of Extinctions*, Interscience Publishers 1966.

24. P. Ginzburg and A. V. Zayats, "Localized Surface Plasmon Resonances in Spatially Dispersive Nano-Objects: Phenomenological Treatise", ACS Nano **7**, 4334–4342 (2013).

25. C.F. Bohrenand D. R. Huffman, Absorption *and Scattering of Light by Small Particles*, Wiley-VCH, 1998.

26. R. C. Rumpf and J. J. Pazos, "Optimization of planar self-collimating photonic crystals," J. Opt. Soc. Am. A **30**, 1297-1304 (2013).

27. P. Ginzburg, A. V. Krasavin, A. N. Poddubny, P. A. Belov, Y. S. Kivshar, and A. V. Zayats, "Self-Induced Torque in Hyperbolic Metamaterials", Phys. Rev. Lett. 111, 036804 (2013).

28. http://comsol.com/.

29. J. A. Stratton, *Electromagnetic Theory*, Swedenborg Press, 2010.

30. A. Potemkin, A.N. Poddubny, P.A. Belov, Yu.S. Kivshar, "Green's function in hyperbolic medium", Phys. Rev. A **86**, 023848 , (2012).

31. P. Fan, U. Chettiar, L. Cao, F. Afshinmanesh, N. Engheta, and M. Brongersma, "An Invisible Metal-Semiconductor Photodetector," Nature Photon. **6**(6), 380-385 (2012).